\documentclass[final,5p,times,twocolumn,authoryear]{elsarticle}

\usepackage{graphicx}
\usepackage{amssymb}
\usepackage{listings}
\usepackage{hyperref}
\usepackage{nameref}
\usepackage{rotating}
\usepackage{hyperref}

\journal{Astronomy and Computing}

\begin{document}

\begin{frontmatter}

\title{Corral Framework: Trustworthy and Fully Functional Data Intensive Parallel Astronomical Pipelines}

  \author[iate,fceia]{J. B. Cabral}\corref{mycorrespondingauthor}
      \cortext[mycorrespondingauthor]{Corresponding author}
      \ead{jbcabral@oac.unc.edu.ar}

  \author[iate]{B. S\'anchez}
  \author[utrgv,utsa]{M. Beroiz}
  \author[iate]{M. Dom\'{i}nguez}
  \author[iate]{M. Lares}
  \author[iate]{S. Gurovich}
  \author[cifasis]{P. Granitto}

\address[iate]{
   Instituto De Astronom\'ia Te\'orica y Experimental -
   Observatorio Astron\'omico C\'ordoba (IATE--OAC--UNC--CONICET),
   Laprida 854, X5000BGR, C\'ordoba, Argentina.}
\address[fceia]{
   Facultad de Ciencias Exactas, Ingenier\'{i}a y Agrimensura, UNR,
   Pellegrini 250 - S2000BTP, Rosario, Argentina.}
\address[utrgv]{
   University of Texas Rio Grande Valley (UTRGV),
   One West University Blvd.
   Brownsville, Texas 78520, USA.}
\address[utsa]{
   University of Texas at San Antonio (UTSA),
   1 UTSA Circle, San Antonio, TX 78249, USA.}
\address[cifasis]{
   Centro Internacional Franco Argentino de Ciencias de la
   Informaci\'on y de Sistemas (CIFASIS, CONICET--UNR),
   Ocampo y Esmeralda, S2000EZP,
   Rosario, Argentina.}

\begin{abstract}
Data processing pipelines represent an important slice of 
the astronomical software library that include chains of processes 
that transform raw data into valuable information via data reduction and analysis.
In this work we present Corral, a Python framework for astronomical pipeline generation.
Corral features a Model-View-Controller design pattern on top of an SQL Relational Database
capable of handling: custom data models; processing stages; and communication alerts,
and also provides automatic quality and structural metrics based on unit testing.
The Model-View-Controller provides concept separation between the user logic and
the data models, delivering at the same time multi-processing
and distributed computing capabilities.
Corral represents an improvement over commonly found data processing
pipelines in Astronomy since the design pattern eases the programmer from dealing with processing flow and parallelization issues,
allowing them to focus on the specific algorithms needed for the successive data transformations and at the same time provides a broad measure of quality over the created pipeline.
Corral and working examples of pipelines that use it are available to the community
at \url{https://github.com/toros-astro}.
\end{abstract}


\end{frontmatter}
\section{Introduction}
\label{section:intro}

The development of modern ground--based and space--born telescopes, covering
all observable windows in the electromagnetic spectrum, and an ever increasing
variability interest via time--domain astronomy have raised the necessity for
large databases of astronomical observations.
The amount of data to be processed has been steadily increasing,
imposing higher demands over: quality; storage needs and analysis
tools.
This phenomenon is a manifestation of the deep transformation that Astronomy is going through,
along with the development of new technologies in the Big Data era.
In this context,
new automatic data analysis techniques have emerged as the preferred solution
to the so-called ``data tsunami'' \citep{cavuoti_data-rich_2013}.

The development of an information processing pipeline is a natural consequence of
science projects involving the acquisition of data and its posterior analysis.
Some examples of these data intensive projects include The Dark Energy Survey Data Management System \citep{mohr2008dark},
designed to exploit a camera with 74 CCDs at the Blanco telescope
to study the nature of cosmic acceleration;
the Infrared Processing and Analysis Center \citep{masci2016ipac},
a near real-time transient-source discovery engine for the
intermediate Palomar Transient Factory \citep[iPTF][]{kulkarni2013intermediate}; and the Pan-STARRS PS1 Image Processing Pipeline \citep{magnier2006pan},
performing the image processing and analysis for
the Pan-STARRS PS1 prototype telescope data
and making the results available to other systems within Pan-STARRS and Vista survey pipeline that includes VIRCAM, a 16 CCD nearIR camera for the VISTA Data flow system \cite{emerson_vista_2004} .
In fact, the implementation of pipelines in Astronomy is a common task
to the construction of surveys \citep[e.g.][]{marx_prototype_2015, hughes_2016, hadjiyska_2012},
and it is even used to operate telescopes remotely, as described in \citet{kubanek_petr_rts2_2010}.
Standard tools for pipeline generation have already been developed and can be found in the literature.
Some examples are Luigi\footnote{Luigi: \url{https://luigi.readthedocs.io/}},
which implements a method for the creation of distributive pipelines;
OPUS \citep{rose1995opus}, conceived by the
\textit{Space Telescope Science Institute}; and more recently
Kira \citep{zhang2016kira}, a distributed tool focused on astronomical
image analysis.
In the experimental sciences, collecting, pre-processing and storing data
are common recurring patterns regardless of the science field or the nature of the experiment.
This means that pipelines are in some sense re-written repeatedly.
%
A more efficient approach would exploit existing resources to build new tools and perform new tasks,
taking advantage of established procedures that have been widely tested by the community.
Some successful examples of this are the \textit{BLAS} library for Linear Algebra, the
package \textit{NumPy} for Python \citep{van2011numpy} and the random number generators.
Modern Astronomy presents plenty of examples where pipeline development is crucial.
In this work, we present a python framework for astronomical pipeline generation developed
in the context of the TOROS collaboration
\citep[``Transient Optical Robotic Observatory of the South",][]{toros_diaz_2014}.
The TOROS project is
dedicated to the search of electromagnetic counterparts to gravitational wave (GW) events, as a response
to the dawn of Gravitational Wave Astronomy.
TOROS participated in the first observation run O1 of the Advanced LIGO GW interferometer \citep{ligo_seminal, O1discoveries}
from September 2015 through January 2016 with promising results \citep{resultsToros_beroiz_2016} and is currently attempting to deploy a wide-field optical telescope in Cord\'on Mac\'on, in the
Atacama Plateau, northwestern Argentina \citep{renzi2009caracterizacion, tremblin2012worldwide}.
The collaboration faced the challenge of starting a robotic observatory in
the extreme environment of the Cord\'on Mac\'on.
Given the isolation of this geographic location
(the site is at 4,600~m~AMSL and the closest city is 300 km away), human interaction
and that Internet connectivity is not readily available,
this imposes strong demands for in-situ pipeline data processing and storage requirements
along with failure tolerance issues.
To assess this, we provide the formalization of a pipeline framework based on the
well known design pattern \textit{Model--View--Controller} (MVC), and
an Open Source BSD-3 License\footnote{BSD-3 License: \url{https://opensource.org/licenses/BSD-3-Clause}}
pure Python package capable of creating a high performance abstraction layer
over a data warehouse, with multiprocessing data manipulation and quality assurance reporting.
This provides simple Object Oriented structures that seizes the power of modern multi-core
computer hardware.
On the assurance reporting given the massive amount of data expected to be processed,
Corral extracts quality assurance metrics for the pipeline run, useful for error debugging.

This work is organized as follows.
In \autoref{section:pipearch} the pipeline formalism and the
relevance of this architecture is discussed, in
\autoref{section:framework} the framework and the design choices
made are explained.
The theoretical ideas are implemented into code as an Open Source Python software tool, as shown in
\autoref{section:results}.
In \autoref{case:code_examples} a short introductory code case for Corral is shown,
followed by \autoref{section:under_hood} where a detailed explanation of the internal
framework's mechanisms and their overhead is discussed, and in \autoref{section:comparison}
a comparison between Corral and other similar projects is shown. Finally in
\autoref{section:real_pipelines} three  production-level pipelines,
each built on top of Corral are listed.
In \autoref{section:conclusions} conclusions, discussion and future highlights of the project
can be found.
Finally the appendix \hyperref[appendixa]{appendix A} presents two brief
examples about experiences of pipeline development with Corral,
and \hyperref[appendixb]{appendix B} where a table that compares Corral with
other pipeline framework alternatives can be found.
%
\section{Astronomical Pipelines}
\label{section:pipearch}
Typical pipeline architecture involve chains of processes that consume a data flow, such
that every processing stage is dependent output of a previous stage.
According to \citet{bowman-amuah_processing_2004}, any pipeline formalism must include
the following entities:
\begin{description}
  \item \textbf{Stream}:  The \textit{data stream} usually means a
     continuous flow of data produced by an experiment that needs to
     be transformed and stored.
  \item \textbf{Filters}: a point where an atomic action is being
     executed on the data and can be summarized as stages of the
     pipeline where the data stream undergoes transformations.
  \item \textbf{Connectors}: the bridges between two filters.  Several
     connectors can converge to a single filter, linking one stage of
     processing with one or more previous stages.
  \item \textbf{Branches}: data in the stream may be of a different
     nature and serve different purposes, meaning that pipelines
     can host groups of filters on which every kind of data must pass,
     as well as a disjoint set of filters specific to different
     kinds of data. This concept allows pipelines the ability to
     process data in parallel whenever data is independent.
\end{description}
This architecture is commonly used on experimental projects that need
to handle massive amounts of data.
We argue that it is suitable for managing the data flow
from telescopes immediately after data ingestion through to the
data analysis.

In general, most dedicated telescopes or observatories have at
least one pipeline in charge of capturing, transforming and storing
data to be analyzed in the future, manually or
automatically \citep{klaus_kepler_2010, tucker_sloan_2006,
emerson_vista_2004}.
This is also important because many of the upcoming large astronomical
surveys
\citep[e.g. \href{https://www.lsst.org/}{LSST}, ][]{lsst_2008}, are
expected to be in the PetaByte scale in terms of raw
data,\footnote{LSST System \& Survey Key Numbers: \url{https://www.lsst.org/scientists/keynumbers}},
\footnote{LSST Petascale R\&D Challenges: \url{https://www.lsst.org/about/dm/petascale}}, meaning that a faster
and more reliable type of pipeline engine is needed. LSST is currently
maintaining their own foundation for pipelines and data management software
\citep{axelrod2010open}.

\section{Framework}
\label{section:framework}

Most large projects in the software industry start from a common baseline
defined by an already existent framework.

The main idea behind a framework is to offer a theoretical
methodology that significantly reduces the repetition of code,
allowing the developer to extend an already existent functionality, optimizing time, costs and other resources \citep{baumer1997framework, pierro_web2py_2011}.
A framework also offers its own flow control and rules to write extensions in a common way,
which also facilitates the maintainability of the code.

\subsection{The Model-View-Controller (MVC) pattern}
\label{mvc}

MVC is a formalism originally designed to define software with visual interfaces
around a data driven (DD) environment \citep{krasner_description_1988}.
The MVC approach was successfully adopted by most
modern full stack web frameworks including 
Django\footnote{Django: \url{https://www.djangoproject.com/}}, 
Ruby on Rails\footnote{Ruby on Rails: \url{https://rubyonrails.org}}, and others.
The main idea behind this pattern is to split the problem into three main parts:
\begin{itemize}
 \item the \textit{model} part, which defines the \textbf{logical} structure of the data,
 \item the \textit{controllers}, which define the logic for \textbf{accessing} and \textbf{transforming} the data for the user, and
 \item the \textit{view}, in charge of \textbf{presenting} to the user the data stored in the model, managed by the controller.
\end{itemize}
In general these three parts were initally defined by the \textit{Object Oriented
Paradigm} \citep[OOP,][]{coad1992object}.
MVC implementation provides self contained, well defined,
reusable and less redundant modules, all of these key features
are therefore necesarry for any big collaborative software project such as astronomical data pipeline.


\subsection{Multi-processing and pipeline parallelization}

As stated before, pipelines can be branched when the chain of data processing
splits into several independent tasks.
This can be easily exploited so that the pipeline takes full advantage of the available
resources that a multi-core machine provides.
Furthermore with this approach, the distribution of tasks inside a network environment,
such as in a modern cluster is simple and straightforward.
%
\subsection{Code Quality Assurance}
\label{qa}

Software quality has become a key component to software development.
According to \citet{feigenbaum1983total},
%
\\
\textit{"Quality is a \textbf{customer} determination, not an engineer's determination,
not a marketing determination, nor a general management determination.
It is based on the customer's actual experience with the product or service,
measured against his or her requirements --
stated or unstated, conscious or merely sensed,
technically operational or entirely subjective --
and always representing a moving target in a competitive market".}\\

In our context, a \textit{customer} is not a single person but a
\textit{role} that our scientific requirements define, and the \textit{engineers}
are responsible for the design and development of a pipeline able
to satisfy the functionality defined by those requirements.
Measuring the quality of software is a task that involves the
extraction of qualitative and quantitative metrics.

One of the most common ways to measure software quality
is \textit{Code Coverage} (CC).
%
%
CC relies on the idea of \textit{unit-testing}.
The objective of \textbf{unit-testing} is to isolate and show that each part of the program is correct \citep{jazayeri_trends_2007}.
Following this, the CC is the percentage of code executed by the unit tests \citep{miller1963systematic}.

Another interesting metric is related to the maintainability of the software.
Although this may seem a subjective parameter, it can
be measured by using a standardization of code style. The number
of style deviations as a tracer of code maintainability.
It is also interesting to define quality based on hardware--use
and its related performance given the software.
This is commonly known as \textit{profiling}
A software profile aids in finding bottlenecks in the resource utilization of the computer,
such as processor and memory use, I/O devices, or even energy consumption \citep{gorelick2014high}.
The broader profile type classification splits into
\textit{application profiling} and \textit{system profiling}.
Application profiling is restricted to the currently developed
software, while system profiling tests the underlying
system (databases, operating system and hardware) looking for
configuration errors that may cause inefficiencies or overconsumption
\citep{gregg2013systems}.
There are different techniques to obtain this information depending on the
unit of analysis and the data sampling method.
There are profilers that evaluate the application in general
(\textit{application level profiling}), on each function call
(\textit{function level profiling}) or each line of code
(\textit{line profiling}) \citep{gregg2013systems}.
Another profiler classification further divides them into
\textit{deterministic} and \textit{statistic} \citep{roskind2007python}
\citep{schneider_statistical_nodate}.
Deterministic profilers sample data at all times, while statistic
profilers records data at regular intervals, taking note of
the current function and line of execution.
Unlike the analysis unit, which is decided in advance
and is usually modified during the analysis, the choice for a
deterministic profile is based on the need to retrieve precise
measurements, at the cost of speed, since deterministic profiles can slow
down the application execution time by up to a factor of ten.
On the other hand, the statistic profiler method executes the
application at almost true speed.


\section{Results: A Python Framework to Implement reproducible Pipelines
         Based on Models, Steps and Alerts}
\label{section:results}

We designed a Data Driven process based on MVC to generate pipelines for applications in Astronomy that support quality metrics by means of unit-testing and code coverage.
It is composed of several pieces, each one consistent with the functionality
set by traditional Astronomical pipelines and also features
branching option for parallel processing naturally.
This design was implemented on
top of the OO \href{http://python.org}{Python} Language and Relational Databases
in a Open Source BSD-3 licensed software project named \textbf{Corral}\footnote{
Corral Homepage: \url{https://gitub.com/toros-astro/corral}}

\subsection{The Underlying Tools: Python and Relational Databases}
As previously mentioned, Corral is implemented on the \href{http://python.org}{Python}
programming language \footnote{Python Programming Language: \url{http://python.org}};
which has a vast ecosystem of scientific libraries such as NumPy,
SciPy \cite{van2011numpy}, Scikit-Learn \cite{pedregosa2011scikit}, and a
powerful and simple syntax.
Most astronomers are choosing Python as their main tool for data processing, favoring
the existence of libraries such as AstroPy \cite{robitaille2013astropy},
CCDProc\footnote{CCDProc: \url{http://ccdproc.readthedocs.io}}, or
PhotUtils\footnote{PhotUtils: \url{http://photutils.readthedocs.io/en/stable/}}\cite{tollerud2016jwst}, e.t.c.,
It is also worth mentioning that Python hosts a series of libraries for parallelism,
command line interaction tools, and test case design, that are all useful for a smooth
translation from ideas into real working software.

Another key requirement is the storage and retrieval of data in multiple processes,
which led us to use
\textit{Relational Databases Management Systems} (RDBMS).
Relational Databases are a proven standard for data management and have been around for more than thirty years. They support an important number of implementations and it is worth mentioning that amongst the most widely used are Open Source, e.g., PostgreSQL.

\textit{SQL} is a powerful programming language for RDBMS and offers advantages in data consistency, and for search queries.
\textit{SQL} has a broad spectrum of implementations: from smaller,
local applications, accessible from a single process, like SQLite \citep{owens2010sqlite},
to distributed solutions on computer clusters, capable of serving billions of
requests, like Hive \citep{thusoo2009hive}.
This plethora of options allows flexibility in the creation of pipelines,
from personal ones, to pipelines deployed across computing clusters hosting
huge volumes of data and multiple users.
Inside the Python ecosystem, the SQLAlchemy\footnote{SQLAlchemy: \url{http://www.sqlalchemy.org/}} library offers
the possibility of designing model Schema in a rather simple way while at the same time offering enough flexibility so as to not cause dependence issues related to specific SQL dialects. Thus offering a good compromise to satisfy different needs of the developer.

\subsection{From Model-View-Controller to Model-Step-Alert}

\begin{figure}
\centering
\includegraphics[width=0.42\textwidth]{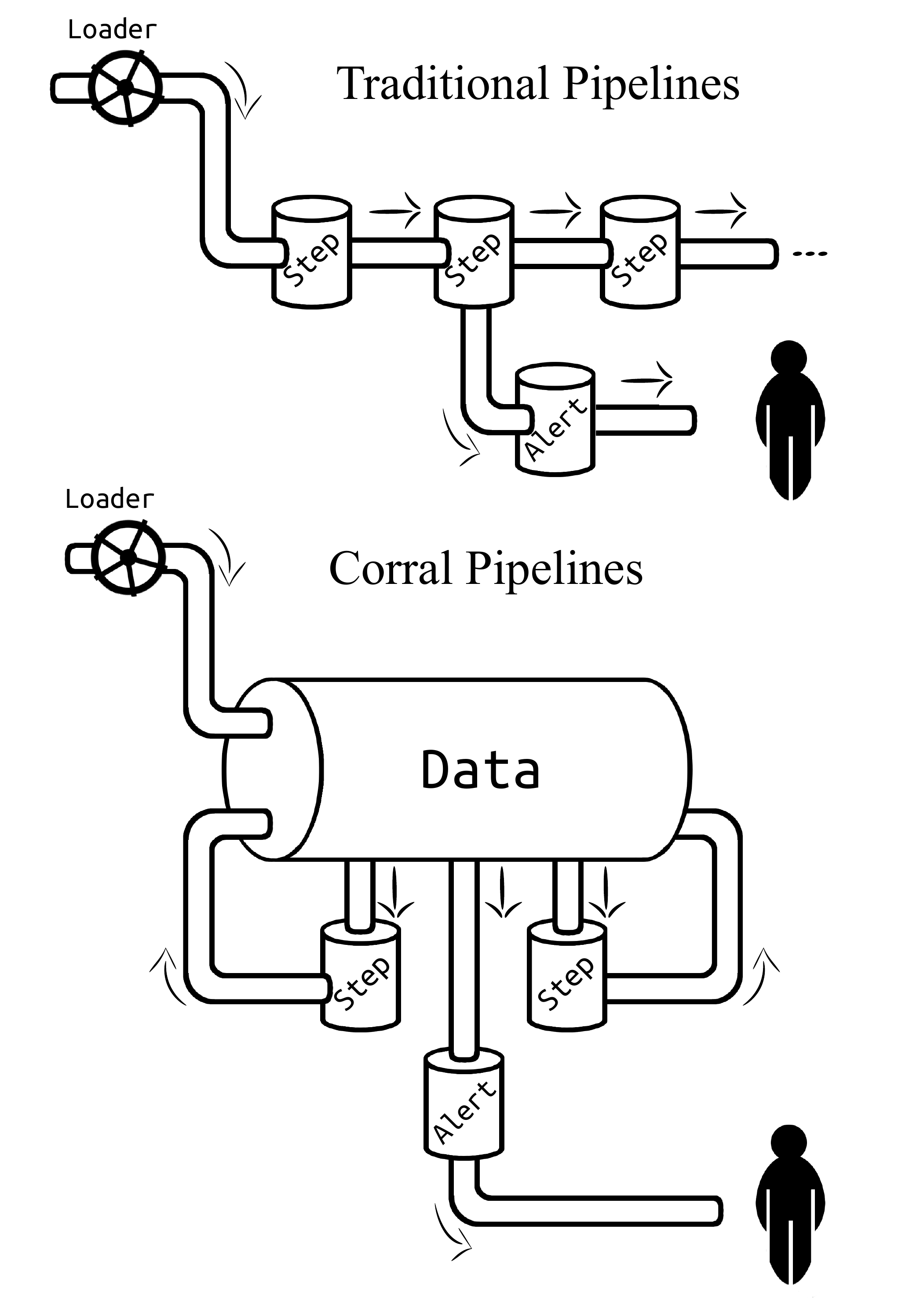}
\caption{\label{fig:concurrent}
    Data road-map in traditional pipeline architecture (upper panel)
    and Corral's pipeline model (lower panel).
    In previous pipeline models data is being processed sequentially
    from one Step to the next one, with the possible scenario of
    parallel branching; while in the model proposed in this work
    Steps are independent entities that interact with a shared
    central data container.
    In both architectures data can be extracted when desired
    conditions are met, and shared at any time to users.
}
\end{figure}

To bridge the gap between traditional MVC terminology and that used to describe pipeline architecture, some terms (e.g. Views and Controllers) have been redefined in this work, to make the code more descriptive for both programmers and scientists.
\begin{description}
\item   \textbf{Models} define protocols for the \textit{stream} of our pipeline. They define data
        structures for the initial information ingest,
        intermediate products and final knowledge of the processing pipeline.
        Since the framework is Data Driven, every stage of the process
        consumes and loads data trough the models that act as the channels of communication amongst the different pipeline components. The models can store data or metadata.
\item   \textbf{Steps} Just like the Models, Steps are defined by classes, and in this case they
        act like \textit{filters} and \textit{connectors}.
        We know mention two different types of steps used by Corral: \textit{loaders}, and \textit{steps}.
        \begin{description}
        \item   \textbf{Loaders} are responsible for feeding the pipeline at its earliest stage. By design choice, a pipeline over Corral can only have one loader.
        \item   \textbf{Steps} select data from the \textit{stream} by imposing constraints,
                and load the data again after some transformation. It follows from this
                that steps are both filters and connectors, that implicitly include branching.
        \end{description}
\item   \textbf{Alerts} To define \textit{views} we take that concepts of Alerts as utilized in some astronomical applications used for follow-up experiments, as inspiration to design the \textit{views}.
        In Corral \textit{views} are called \textit{Alerts} and are special events triggered by a particular state of the data \textit{stream}.
\end{description}
Some Python code examples regarding each of these "pipeline processors" can be found in
\autoref{case:code_examples} and \hyperref[appendixa]{appendix A}
A complete picture of the framework and the user defined elements of the pipeline
along with their interactions,
is displayed in figure \ref{fig:concurrent}.

\subsection{Some Words About Multiprocessing and Distributed Computing}
As previously mentioned our project is built on top of a RDBMS, which by design
can be accessed concurrently from a local or network computer.
Every step accesses only the filtered part of the stream,
so with minimum effort one can deploy the pipeline in one or more
computers and run the steps in parallel.
All the dangers of data corruption are avoided
by the ACID (Atomicity, Consistency, Isolation and Durability)
properties of the database transactions.
%
This approach works well in most scenarios, but there are some unavoidable 
drawbacks, that arise for example in \textit{real-time} processing, where 
consistency is less important than availability.
Corral takes advantage of this technology to start a process for each Loader,
Step or Alert within the pipeline, and allows interaction with the data stored in the database.
If needed, groups of processes can be manually distributed on nodes of a
cluster where the nodes will interact with the database remotely.

It is worth noting that any inter-processes direct communication is forbidden by design,
and the only way to exchange messages is through the database.
On this last particular point, the absolute isolation of the processes,
is guaranteed by the MVC pattern.
%
\subsection{Quality -- Trustworthy Pipelines}
One important requirement for a pipeline is the reliability of its results.
A manual check of the data is virtually impossible when its volume scales to the TeraByte range.
In our approach we suggest a simple unit testing approach to check the status of the stream before and after every
\textit{Step}, \textit{Loader} or \textit{Alert}.

Because tests are unitary, Corral guarantees the isolation of each
test by creating and destroying the stream database before and after execution of the test.
If you feed the stream with a sufficient amount of heterogeneous data you can check most of the pipeline's functionality before the production stage. Finally we provide capabilities to create reports with all the structural and quality assurance information about the pipeline in convenient way, and a small tool to profile
the CPU performance of the project.
\subsubsection{Quality Assurance Index (QAI)}
We recognize the need of a value to quantify the pipeline software quality.
For example, using different estimators for the stability and maintainability of the code, we
arrived at the following Quality Index:
\begin{displaymath}
QAI = \frac{ \Theta \times \Lambda_{Cov} \times R_{PT}}{\gamma}
\end{displaymath}
and $\gamma$ is a penalty factor defined as:
\begin{displaymath}
\gamma = \frac{1}{2} \times \left( 1 + exp \left( \frac{N_{SError}}{\tau \times N_f} \right)\right)
\end{displaymath}
$\Theta$ is 1 if every test passes or 0 if any one fails,
$R_{PT}$ is the ratio of tested processors (\textit{Loader, Steps and Alerts})
to the total number of processors, $\Lambda_{Cov}$ the code coverage (between 0 and 1),
$N_{SError}$ is the number of style errors, $\tau$ is the style tolerance,
and $N_f$ is the number of files in the project.
The number of test passes and failures are the unit-testing results,
that provide a reproducible and upda\-table manner
to decide whether your code is working as expected or not.
The $\Theta$ factor is a critical parameter of the index, since it is discrete,
and if a single unit test fails it will set the QAI to zero, in the spirit that if
your own tests fail then no result is guaranteed to be reproducible.
The $R_{PT}$ factor is a measure of how many of the different processing stages critical to the pipeline
are being tested (a low value of this parameter should be interpreted as a need to
write new tests for each pipeline stage).
The $\Lambda_{Cov}$ factor shows the percentage of code that
is being executed in the sum of every unit test;
this displays the "quality of the testing" (a low value should be interpreted
as a need to write more extensive tests, and it may 
correlate with a low number of \textit{processors} being tested, 
that is a low $R_{PT}$).
$N_{NSerr}/(\tau \times N_f)$ is the scaling factor for the exponential.
It comprises the information regarding style errors, attenuated
by a default or a user-defined tolerance $\tau$ times the number
of files in the project $N_f$. 
The exponential function expresses the fact that a certain number of 
style errors isn't critical, but after some point this seriously compromises
the maintainability of the software project, and in this situation 
$\gamma$ strongly penalizes the quality index.
The factor $1/2$ is a normalization constant, so that $QAI \in [0, 1]$.
This index aims to encode in a single figure of merit how well the pipeline
meets the requirements specified by the user.
We note that this index represents a confidence metric. 
Thus a pipeline could be completely functional even if every test fails, 
or if no tests are yet written for it. 
%
And in the opposite direction, the case where every test passes and the pipeline
is delivering wrong or bogus results is possible.
The $QAI$ index attempts to answer the question of pipeline
reliability and whether a particular pipeline can be trustworthy.
It should not be confused with the pipeline's speed,
capability, or any other performance metric.
\subsubsection{Determining the default error tolerance in a python project}
Corral, as earlier mentioned is written in Python,
which offers a number of third party libraries for
style validation. Code style in Python is standardized by the
PEP 8 document\footnote{PEP8: \url{https://www.python.org/dev/peps/pep-0008}}.
Flake8\footnote{Flake8: \url{http://flake8.pycqa.org/}} is a style tool that allows
the user to measure the number of style errors, that reflects the maintainability of the project.

\begin{figure}
\centering
\includegraphics[width=0.5\textwidth]{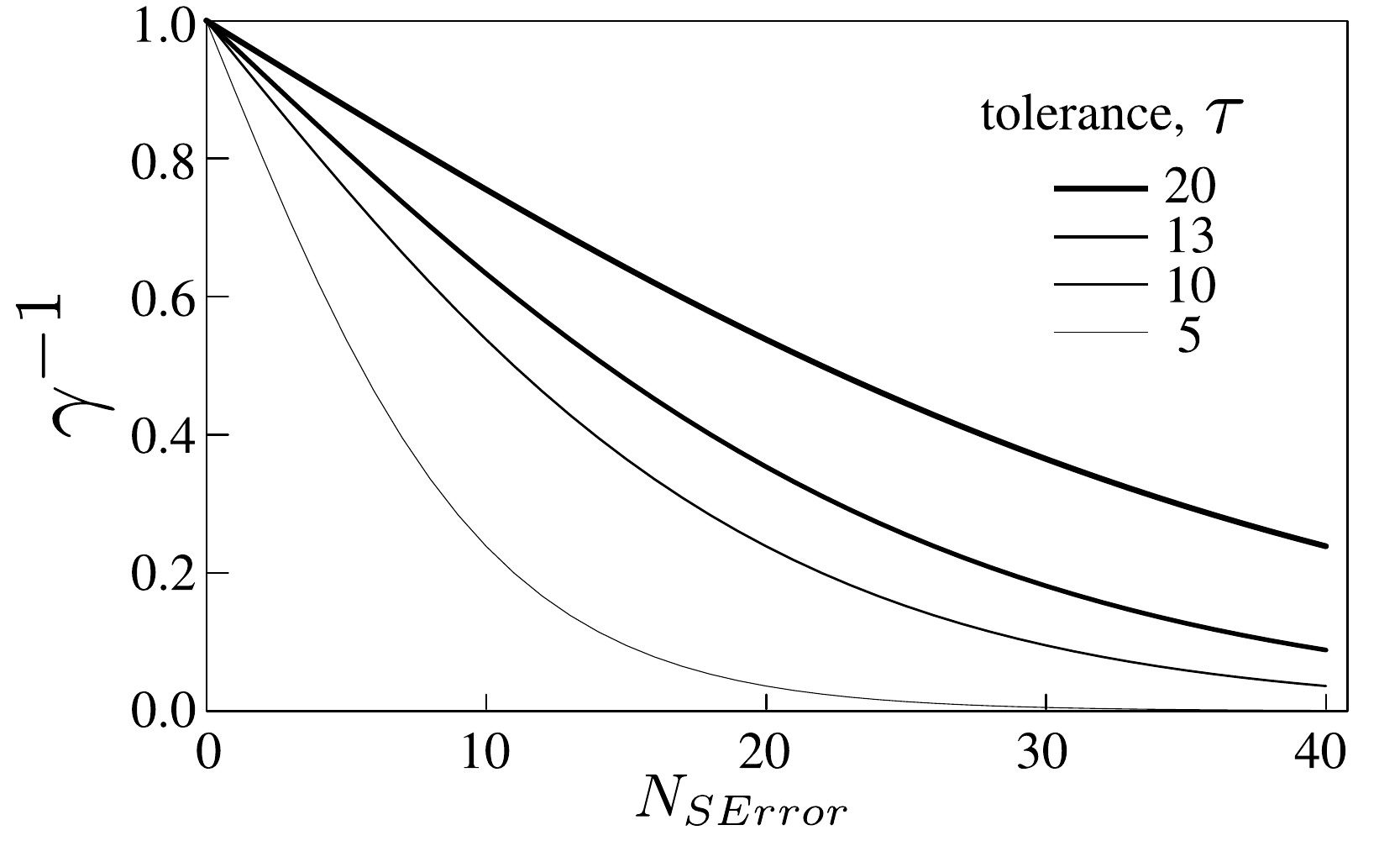}
\caption{\label{fig:tau} Ideal test pass, and coverage QAI curves, with $N_{SErr} \in [0, 40]$ for
    four values of $\tau$, for only one file.
    A damp in the error slope decay can be observed when $\tau$
    is higher.
}
\end{figure}

For the $QAI$ measurement of Corral a key detail was in the determination of the amount of style errors developers tollerate as normal. For this we collected data from nearly $4000$ public Python source code files. The number of style errors was determined using Flake8 and the inter-quartile mean
was determined as a measurement for $\tau$. A $\tau$ of $\sim 13$ was found. It is important to note that this value can be
overridden by the user if a stronger QAI is required (Fig. \ref{fig:tau}).
\subsubsection{Quality and Structural reporting}
Corral includes the ability to automatically
generate documentation, creating a manual and quality reports in
Markdown\footnote{Markdown: \url{https://daringfireball.net/projects/markdown/}}
syntax that can be easily converted to a preferred format such as
PDF \citep{iso2005document}, LaTeX \citep{lamport1994latex} or HTML \citep{priceiso}.
A Class-Diagram \citep{booch2006uml} generation tool for the
Models defined using Graphviz\footnote{Graphviz: \url{http://www.graphviz.org/}}
are also available for the pipelines created with Corral.
We provide in \hyperref[appendix_exo]{Appendix A.4} an example of a pipeline (EXO) with an implementation of tests and the corresponding results of the QA report.

\subsubsection{Profiling}
Corral offers a deterministic tool for the performance analysis
of CPU usage at a function level\footnote{Function Level CPU Profiling:
shows function call times and frequency, as
well as the chain of calls they were part of based on the receiver of the call.}
during the execution of unit tests.
It is worth noting that in case a pipeline shows signs of lagging or
slowing down, running a profiler over a unit test session can help locate bottlenecks.
However for rigurous profiling, real data on real application runs should
be used, instead of unit testing.

Another kind of profiling at the application level could be carried out
manually using existing Python ecosystem tools such as \textit{memory\_profiler}\footnote{memory\_profiler:\url{https://pypi.python.org/pypi/memory_profiler}}
\footnote{memory\_profiler:\url{https://pypi.python.org/pypi/memory_profiler}}
(memory line level deterministic profiling),
\textit{statsprof.py}\footnote{statsprof.py:\url{https://github.com/smarkets/statprof}}
(statistic function level profiling), o
\textit{line\_profiler}\footnote{line\_profiler: \url{https://github.com/rkern/line_profiler}}
(line level deterministic) amongst other tools.
We note that although some application level profiling tools are included or suggested, Corral was never intended to offer a system profiling tool. Nor does it claim to offer data base profiling, or I/O, energy profiling, network profiling, etc.
\subsubsection{Final words about Corral quality}
The framework does not contain any error backtrace concept, or retry attempts in
processing. Each processor should be able to handle correctly the required information on its conditions.
It is implicitly expected that the quality tools offered serve to unveil code errors as well.

If the pipeline's developer achieves a high code coverage and is able to test enough data
diversification, the possible software bugs can decrease substantially, up to 80\% \citep{jeffries2007guest}.
%
\section{Study case: A Simple Pipeline to Transform (x, y) Image Coordinates to world coordinates (RA, Dec)}
\label{case:code_examples}
A few examples are given below to illustrate each part of a toy model pipeline
built over Corral.
A more extended example can be found in \hyperref[appendixa]{appendix A} and in the
TOROS GitHub repository page \url{https://github.com/toros-astro/toritos}, where a fully functional astronomical
image preprocessing pipeline is available.

We encourage the interested users to read the full Corral tutorial located at:
\url{http://corral.readthedocs.io}
\subsection{A \textit{Model} example}
In the following example a Model for an astronomical source is shown. It uses
both $(x,y)$ and $(RA, Dec)$ coordinates and a field given for apparent magnitude.
The identification field is automatically settled.
%
\begin{verbatim}
# this code is inside mypipeline/models.py
from corral import db

class PointSources(db.Model):
  "Model for star sources"
  __tablename__ = "PointSources"

  id = db.Column(
    db.Integer, primary_key=True)
  x = db.Column(db.Float, nullable=False)
  y = db.Column(db.Float, nullable=False)
  ra = db.Column(db.Float, nullable=True)
  dec = db.Column(db.Float, nullable=True)

  app_mag = db.Column(db.Float, nullable=False)
\end{verbatim}
%
%
\subsection{A \textit{Loader} example}
In the following example a Loader is shown, where the PointSource
Model from above example is filled with new data. We note that the $(RA, Dec)$ fields in PointSource Model are allowed to be null, and therefore there is no need to set them a priori.
%
\begin{verbatim}
# this code is inside mypipeline/load.py
from astropy.io import ascii
from corral import run

from mypipeline import models

class Load(run.Loader):
  "Pipeline Loader"
  def setup(self):
    self.cat = ascii.read(
        'point_sources_cat.dat')

  def generate(self):
    for source in self.cat:
      src = models.PointSource()
      src.x = source['x']
      src.y = source['y']
      src.app_mag = source['app_mag'])
      yield src
\end{verbatim}
%
\subsection{A \textit{Step} example}
An example Step is displayed below, where a simple data transformation
is conducted.
Where the step takes a set of sources and transforms their $(x,y)$ coordinates to $(RA, Dec)$:
lines 10-12 show definitions for Class-level attributes,
which are responsible for this query. The framework retrieves the data
and serves it to the \texttt{process} method (line 14), which
executes the relevant transformation and loads the data for each
source.
%
%
\begin{verbatim}
# this code is inside mypipeline/step.py
from corral import run
from mypipeline import models

import convert_coord

class StepTransform(run.Step):
  "Step to transform from x,y to ra,dec"

  model = models.PointSource
  conditions = [model.ra == None,
                model.dec == None]

  def process(self, source):
    x = source.x
    y = source.y

    source.ra, source.dec = convert_coord(x, y)
\end{verbatim}
%
%
\subsection{\texttt{Alert} Example}
In the example below an Alert is triggered when a
state satisfies a particular condition of data in the stream. A group of communication channels are activated.
In this particular case an email and an Astronomical Telegram
\citep{rutledge1998astronomer} are posted
whenever a point source is detected in the vicinity of $\textrm{Sgr A}^{*}$,
near the center of the galaxy.
%
\begin{verbatim}
# this code is inside mypipeline/alert.py
from corral import run
from corral.run import endpoints as ep

from mypipeline import models

class AlertNearSgrA(run.Alert):

  model = models.PointSource
  conditions = [
    model.ra.between(266.4, 266.41),
    model.dec.between(-29.007, -29.008)]
  alert_to = [ep.Email(["sci1@sci.edu",
                        "sci2@sci.edu"]),
              ep.ATel()]
\end{verbatim}
%
%
\subsection{Running your Pipeline}
As seen in the previous sections we define our protocol for the stream and
the actions to be performed in order to covert the coordinates; but code is never defined
to schedule the execution. Nevertheless applying the Corral MVC equivalent pattern
guarantees that every loader and step is to be executed independently which means that both tasks
run as long as there is data to work with.
In every Step/Loader/Alert there is a guarantee of data consistency since each is attached to an SQLAlchemy \texttt{session}, and every task is linked to a database transaction. Since every SQL transaction is atomic, so
the process is executed or fails, thus lowering the risk of data corruption in the stream.
\newpage
Typically for a given pipeline with defined: Loader; Alert and Steps, the following output is produced:
%
\begin{verbatim}
$ python in_corral.py run-all
[mypipeline-INFO@2017-01-12 18:32:54,850]
    Executing Loader
    '<class 'mypipeline.load.Load'>'
[mypipeline-INFO@2017-01-12 18:32:54,862]
    Executing Alert
    '<class 'mypipeline.alert.AlertNearSgrA'>'
[mypipeline-INFO@2017-01-12 18:32:54,874]
    Executing Step
    '<class 'mypipeline.load.StepTransform'>'
[mypipeline-INFO@2017-01-12 18:33:24,158]
    Done Alert
    '<class 'mypipeline.alert.AlertNearSgrA'>'
[mypipeline-INFO@2017-01-12 18:34:57,158]
    Done Step
    '<class 'mypipeline.load.StepTransform'>'
[mypipeline-INFO@2017-01-12 18:36:12,665]
    Done Loader
    '<class 'mypipeline.load.Loader'>' #1
\end{verbatim}
%
It can be seen in the time stamps of the executions and task completions for the
Loader, Alert and Steps; that there is no relevant ordering between them.
\subsection{Checking The Pipeline Quality}
\subsubsection{Unit Testing}
Following the cited guidelines of \cite{feigenbaum1983total} who states
that quality is a user defined specification agreement, it is necessary to make
this explicit in kind in code that for Corral is achieved in a unit-test.
Below an example is included showing a basic test case for our Step. The Corral test,
feeds the stream with some user defined mock data, then
runs the Step and finally checks if the result status of the stream meets
the expected value.
%
\begin{verbatim}
# this code is inside mypipeline/tests.py
from corral import qa

from mypipeline import models, steps

import convert_coord

class TestTransform(qa.TestCase):
  "Test the StepTransform step"
  subject = steps.StepTransform

  def setup(self):
    src = models.PointSource(
      x=0, y=0, app_mag=0)
    self.ra, self.dec = convert_coord(0, 0)
    self.save(src)

  def validate(self):
    self.assertStreamHas(
      models.PointSource,
      models.PointSource.ra==self.ra,
      models.PointSource.dec==self.dec)
    self.assertStreamCount(
      1, models.PointSource)
\end{verbatim}
%
As shown in lines 12-16 the \texttt{setup()} method is in charge of
creating new data --whose transformed result is already known--,
so then \texttt{validate()} asserts the outcome of \texttt{StepTransform}.
This process would be repeated if more tests were defined, and
an important caveat is that, mocked data streams are private to each unit test
so will never collide producing unexpected results.
\subsubsection{Quality Report and Profiling}
Corral provides built-in functionalities to communicate quality assurance information:
\begin{enumerate}
\item   \texttt{create-doc}: This command generates a Markdown
            version of an automatically generated manual for the pipeline. It includes information on
            Models, Loader, Steps, Alerts, and command line interface
            utilities, using docstrings from the code itself.
\item   \texttt{create-models-diagram}: This creates a Models Class Diagram
            \citep{booch2006uml}
            in Graphviz dot format \citep{ellson2001graphviz}.
\item   \texttt{qareport}: Runs every test and Code Coverage evaluation,
            and uses this to create a Markdown document detailing the
            particular results of each testing stage, and finally calculates
            the QAI index outcome.
\item  \texttt{profile}: Executes all existing tests and deploys an interactive web interface to evaluate the performance of different parts of the pipeline.
\end{enumerate}

With these four commands the user can get a detailed report about structural
status, as well a global measurement of quality level of the pipeline.
%
\section{Corral: Under the hood}
\label{section:under_hood}
Put simply, Corral is a Pipeline environment that autoconfigures itself on each user command for which the following operations are exectued:
\begin{enumerate}
    \item First of all \textit{in\_corral.py} inserts the path of \textit{settings.py}
    into the  environment variable \texttt{Corral\_SETTINGS\_MODULE}.
    With this simple action every Corral component knows where to find
    the pipeline's configuration.
    \item The command line parser is created, and commands
    provided by Corral are made available.
    \item If the user asks for help (with the \texttt{--help|-h} flag)
    --or the requested command does not exist--, the help page is printed to screen.
    \item Given a valid command, the line arguments are parsed so the requested action
    is identified, and set to be executed.
    \item Based on the requested command, the framework would work in three DBMS modes:
        \begin{description}
            \item \textbf{Mode in}: The production DB is configured as the
            back-end to each model. This kind of configuration is used since in general commands require knowledge of the data stored in the stream.

            \item \textbf{Mode test}: The test DB is configured (by default in-memory DB is used).
            This mode is used by commands that require destructive operations over the database, eg \texttt{test,coverage,profile}
            \item \textbf{Mode out}: No database is configured. Some commands do not require any operations over the stream, like \texttt{lsstep}, used to list all the existing steps.
        \end{description}
     \item The command true logic is executed.
     \item The \textit{teardown} is executed for every connection.
     \item Finally the python interpreter is shut down
     and the current error code is delivered to the operative system.
\end{enumerate}

As mentioned the most basic functionality of Corral is to find files based in
only one environment configuration. This brings to the developer clear rules to split the functionality in well defined and focused modules.

According to measured estimates, typically 1.5--2 s of overhead is required for any
executed command, depending on the number of processes spawned. Almost 100\% of the overhead is spent before the true command logic is executed.
An interesting point to this is that the running mode of commands strongly affects
the execution time ('out' mode is much faster than 'test' with
in-memory database; and test is much faster than 'in' mode).
%
Other external factors causing potential bottlenecks that are worth mentioning
are database location, I/O, Operating System configuration, or any hardware problems;
although these should produce lack of performance not only to Corral but to
any piece of software the user is executing.
%
%
\section{Comparison with other pipeline frameworks}
\label{section:comparison}
The two main differences between Corral and other similar projects are now explained.
(A comparison of other alternatives is shown in Appendix B):
First, to our knowledge Corral is the first implementations of a pipeline framework that uses MVC; and second the quality integration metrics that give an indication of the trustworthiness of the resulting pipeline.

The use of the MVC design standard imposes the following processing tasks (Loader, Steps and Alerts) result in a strong isolation condition:
every processing stage only interacts with filtered data according to specific criteria  with no bearing on information of the source or data destination.
A major advantage of isolation is the natural parallelization of processing tasks since
no "race condition" is created\footnote{Race Condition: Is a software behavioral term that refers to the mutual and competing need of components to work with data.
This leads to errors when the processing order is not as expected by the programmer.}.
Regarding real-time processing this pattern can be inconvenient, since
batches of data are processed asynchronously, leading to random ordering
of data processing and writing onto the Database.

Corral features "integration quality" utilities as
an important tool set that builds confidence on the pipeline.
This works when unit-tests are available for the current
pipeline and in these cases Corral can automatically generate
reports and metrics to quantify reliability.
Corral was designed to optimize pipeline confidence in terms of
some global notion of quality, which implies revision of data in each
processing stage.
Pelican \footnote{Pelican:
\url{http://www.oerc.ox.ac.uk/~ska/pelican/1.0/doc/user/html/user_reference_pipelines.html}}
is the only project integrating tools for pipeline testing, but does not include
extra functionalities based on this concept.
In many other aspects as depicted in \hyperref[appendixb]{appendix B},
Corral is similar to other alternatives. A majority of these alternative also use Python
as the programing language mainly in order to make use of
its vast libraries for data access and manipulation, across multiple formats.

%
\section{Real Pipelines Implemented on Corral}
\label{section:real_pipelines}
To date three pipelines have been implemented on Corral.
\begin{enumerate}
\item The \textbf{TORITOS pipeline} for image pre-processing.
TORITOS is a pilot project
for TOROS which employs a 16" telescope to take images in
the Mac\'on ridge. The pipeline is available at
\url{https://github.com/toros-astro/toritos}.
\item \textbf{Carpyncho} is a Machine Learning facility, for the VVV
\citep{minniti2010vista} survey data,
specifically built to find variable stars in the Galactic Bulge and adjacent disk zone
\citep{cabral2016generacion}.
\item A pipeline for synthetic data generation for
machine learning training, focused on transient detection on survey images
(S\'anchez et al. 2017 in preparation).
\end{enumerate}
%
%
\section{Conclusions and Future Work}\label{section:conclusions}
In this work a pipeline framework is presented that facilitates designing a parallel work flow for data multi processing.
MVC design pattern was employed, that delivers a set of processing entities --Models, Steps, and Alerts-- capable of carrying out a wide variety of scientific data pipeline tasks inside a concurrent scheduled environment.
Last but not least, detailed quality and structural
reports can be extracted and compared to the user's
predefined level of agreement to determine the pipeline
trustworthiness and ultimately the validity of processed data.

Future work includes improvements on Corral's performance
by integrating the framework scheduler over distributed
computing systems.
This could run, for example, on top of 
Apache Hadoop\footnote{Apache Hadoop: \url{http://hadoop.apache.org/}}, or 
Apache Spark\footnote{Apache Spark: \url{https://spark.apache.org/}},
as these are the \textit{state of the art} regarding
data processing capability.
Another possibility for the future is to replace the task scheduler
that currently uses the module \textit{multiprocessing}\footnote{Python multiprocessing:
\url{http://docs.python.org/3/library/multiprocessing.html}}
from the standard Python library, with \textit{Apache storm}
\footnote{Apache Storm: \url{http://storm.apache.org/}}

Some other projects that could make use of this framework include
Weak lensing analysis pipelines, Radio Telescope image
generation and analysis, spectroscopic data analysis on star clusters
and many more.

The capabilities of the presented framework can
be a game-changer in isolated environments where
hardware is operated remotely and resilience is
an important requirement; and since it is
straightforward, Corral  can be used in a wide
variety science cases.
Anyone or any group interested in using Corral is
invited to direct any questions, suggestions,
feature request or advice at
\url{https://groups.google.com/forum/#!forum/corral-users-forum}.

\section{Acknowledgments}
The authors would like to thank to their families and friends, the TOROS collaboration
members for their ideas, and also IATE
astronomers for useful comments and suggestions.

This work was partially supported by the Consejo Nacional
de Investigaciones Cient\'ificas y T\'ecnicas (CONICET, Argentina)
and the Secretaría de Ciencia y Tecnolog\'ia de la Universidad
Nacional de C\'ordoba (SeCyT-UNC, Argentina).
B.O.S., and J.B.C, are supported by a fellowship from CONICET.
Some processing was acheived with Argentine VO (NOVA) infrustructure,
for which the authors express their gratitude.

This research has made use of the
http://adsabs.harvard.edu/, Cornell University xxx.arxiv.org repository,
the Python programming language and SQLAlchemy library,
other packages used can be found at
Corral gihtub webpage.

%
\section*{References}
\label{biblio}
\bibliographystyle{model2-names-astronomy}
\bibliography{corral}


\appendix

Here we present two examples of pipeline development.

\section{A quick--start guide to creating a Corral pipeline}
\label{appendixa}

\subsection{Installation}

The recommended installation method for getting Corral running
is using pip:

\begin{verbatim}
$ pip install -U corral-pipeline
\end{verbatim}

Other methods are also possible, and are detailed in the online
documentation\footnote{http://corral.readthedocs.io/en/latest/install.html}.

\subsection{Simple pipeline examples}

Here we present a quick start guide to use Corral to set up a
pipeline, with two simple examples.

\begin{figure}
\includegraphics[width=0.35\textwidth]{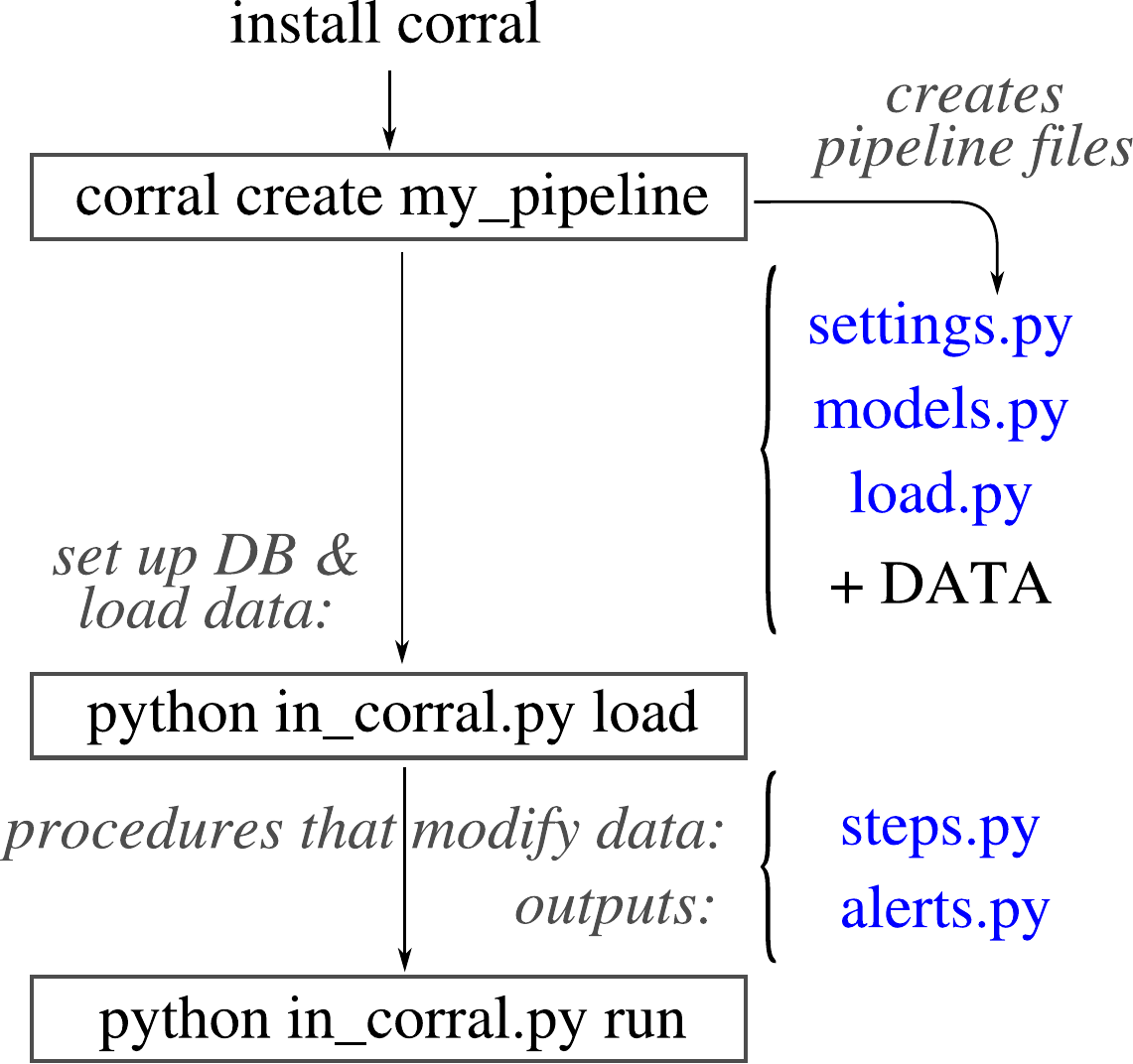}
\caption{Basic scheme of the workflow to build a pipeline using \textsc{Corral}}
\label{fig:scheme}
\end{figure}

The workflow for creating a simple pipeline can be summarized as follows (see also Fig. \ref{fig:scheme}):

\begin{itemize}
   \item Create the pipeline
   \item Define the models
   \item Create the database for the data
   \item Load the data
   \item Define the steps
   \item Run the pipeline
\end{itemize}

We show in what follows two simple examples of pipelines along with
some relevant code that show the basics
to get started with using \textsc{Corral}.

The first example uses the IRIS data, and is intended to compute some
simple statistics.
The IRIS flower data set \citep{fisher1936} is a commonly used multivariate
data set that stores data from 3 species of the iris flower
(''Setosa'', ''Virginica'' and ''Versicolor'').
In the following subsection we show how to implement a pipeline that
reads the IRIS data, stores it in a database and perform some simple
statistics.

A second example uses data from Exoplanet Data Explorer, which is an
interactive web service to exploring data from the Exoplanet Orbit
Database \citep{hann2014}, that stores a compilation of
spectroscopic orbital parameters of exoplanets and stellar parameters of their host
stars.
In this example, we will construct a pipeline to store exoplanets data
and perform some simple exploratory analysis.
The pipeline to be created would be as follows:
\begin{itemize}
   \item Download the data from \url{exoplanets.org}
   \item Update the database
   \item Create a scatter plot of planets period vs. mass
\end{itemize}

In both cases, the pipeline can be constructed using the
\textsc{Corral} framework by performing the following operations:

\begin{itemize}
   \item Create the pipeline, set the location of the data file and
      other configuration parameters in settings\\
      \verb|corral create my_pipeline|
   \item Define the models for the data base.  In this case, tell what
      kind of data will be stored in the database, set names, etc.
   \item Create the database for the data.  This is accomplished
      according to the settings and the models.
   \item Load the data: read the CSV table and save it in the
      database.\\
      \verb|python in_corral.py load|
   \item Define the steps: write the code to be applied to the data.
   \item Run the pipeline: this will read the data from the database
      and perform all the steps defined before.\\
      \verb|python in_corral.py run|
   \item Output data can be obtained through Alerts.\\
      \verb|python in_corral.py check-alerts|
\end{itemize}

In what follows we present a step--by--step guide to carry out this
tasks.

After installation, a pipeline can be created by running the command:
%
\begin{verbatim}
   $ corral create my_pipeline
\end{verbatim}
%
which creates a file in\_corral.py and a new directory 'my\_pipeline'.
The file in\_corral.py is the access point to the pipeline, and allows
commands to be executed inside the environment of the pipeline.
The directory my\_pipeline is the root directory for the pipeline, and
the actual Python package for the project.
Its name is the Python package name that can be used to import
anything inside it (e.g. my\_pipeline.models).
The name my\_pipeline can be replaced by IRIS or EXO, for our examples
on IRIS data or exoplanet data, respectively.
There are some files that are created within the project directory,
each of them with a specific purpose within the Corral framework:

\begin{description}
   \item [my\_pipeline/\_\_init\_\_.py] An empty file that tells
      Python that this directory should be considered a Python
      package.
   \item [my\_pipeline/pipeline.py] The suggested file to globally
      configure the pipeline on execution time.
   \item [my\_pipeline/models.py] Contains the entities (or tables)
      that are stored in the pipeline’s database.
   \item [my\_pipeline/load.py] Contains the Loader class. This would
      be the entry point for raw data to the pipeline stream, before
      going through any defined Steps.
   \item [my\_pipeline/steps.py] Contains all pipeline steps.
   \item [my\_pipeline/alerts.py] Contains the definitions of the
      Alerts, that offer custom communication channels to report
      expected results (a email for instance).
   \item [my\_pipeline/commands.py] Used to add custom console commands, specific for the pipeline.
\end{description}

Each time a new pipeline is created, the first step is to perform the
configuration by editing the file \texttt{settings.py}
This file contains the variables that should be set in order to put
the pipeline into work.
In particular, it stores the location of the settings.py file:
\begin{verbatim}
PATH = os.path.abspath(os.path.dirname(__file__))
\end{verbatim}
%
and also stores information about the locations of data files.
For example, we can create the variable IRIS\_PATH in the IRIS
pipeline that stores the
location of the data file containing the IRIS data, iris.csv
\footnote{\url{https://github.com/toros-astro/corral/raw/master/datasets/iris.csv}}
%
\begin{verbatim}
IRIS_PATH = os.path.join(PATH, "iris.csv")
\end{verbatim}
%
This file contains 5 columns, listing the sepal length, sepal width,
petal length, petal width and name of 150 samples if the IRIS flower
from the original dataset (Fisher 1936).

For the exoplanets pipeline, we store the data in the file
exoplanets.csv\footnote{\url{https://github.com/toros-astro/corral/raw/master/datasets/exoplanets.csv}},
which contains the following columns: name, period and mass of the
planet; star--planet separation; distance to the host star; and mass,
radius, effective temperature and metallicity of the star.
This is a subset of data obtained from \url{exoplanets.org}.
For this pipeline, the only modification in the settings.py file, with
respect to the other example, is:
%
\begin{verbatim}
EXO_PATH = os.path.join(PATH, "exoplanets.csv")
\end{verbatim}

Besides this basic configuration, the names of classes containing
loaders, steps and alerts must be listed in the LOADER, STEPS and
ALERTS variables, respectively.
For example, for the IRIS pipeline we create 4 steps:

\begin{verbatim}
STEPS = [
    "pipeline.steps.StatisticsCreator",
    "pipeline.steps.SetosaStatistics",
    "pipeline.steps.VirginicaStatistics",
    "pipeline.steps.VersicolorStatistics"]
\end{verbatim}

\subsection{Processing of the IRIS data}

As an essential part of the MVC pattern, the
next step in preparing the pipeline is to define the models.
The Model determines the structure of the data, and is defined
in the models.py file.

We can create a model as a database which consists on two tables.
The \textit{Flower} table has 4 columns that store the sepal length and
width, and the petal length and width, in the variables sepal\_l,
sepal\_w, petal\_l, and petal\_w, respectively.
The other table stores
the name of the flower (Name).
Each model is a class that inherits from db.model, so that Corral take
these as tables in the database.
Both tables are linked through their primary keys, which are added
automatically when the database is created.

\begin{verbatim}
from corral import db

class Name(db.Model):

    __tablename__ = 'Name'

    id = db.Column(db.Integer, primary_key=True)
    name = db.Column(db.String(50), unique=True)
\end{verbatim}

\begin{verbatim}
class Flower(db.Model):

    __tablename__ = 'Flower'

    id = db.Column(
        db.Integer, primary_key=True)

    name_id = db.Column(
        db.Integer, db.ForeignKey('Name.id'),
        nullable=False)
    name = db.relationship(
        "Name", backref=db.backref("flowers"))
    sepal_l = db.Column(
        db.Float, nullable=False)
    sepal_w = db.Column(
        db.Float, nullable=False)
    petal_l = db.Column(
        db.Float, nullable=False)
    petal_w = db.Column(
        db.Float, nullable=False)
\end{verbatim}

In the IRIS example, in order to compute a basic statistic (the mean)
for each data column, an additional table must be created to store the
results (and any other value that could be required).
This table in the database is linked to the ''flowers'' table by its
primary key, which is generated automatically.
It must be declared into the \verb|models.py| file:

\begin{verbatim}
class Statistics(db.Model):

    __tablename__ = 'Statistics'

    id = db.Column(db.Integer, primary_key=True)
    name_id = db.Column(
        db.Integer, db.ForeignKey('Name.id'),
        nullable=False, unique=True)
    name = db.relationship(
        "flowers", uselist=False,
        backref=db.backref("statistics"))

    mean_sepal_l = db.Column(
        db.Float, nullable=False)
    mean_sepal_w = db.Column(
        db.Float, nullable=False)
    mean_petal_l = db.Column(
        db.Float, nullable=False)
    mean_petal_w = db.Column(
        db.Float, nullable=False)

    def __repr__(self):
        return "<Statistics of '{}'>".format(
            self.name.name)
\end{verbatim}

Once the model is defined, the database is created with:

\begin{verbatim}
   $ python in_corral.py createdb
\end{verbatim}

At this point the database does not contain any data, so the
\textit{loaders} must be used for that purpose.
This is accomplished with the sentence
%
\begin{verbatim}
   $ python in_corral.py load
\end{verbatim}
that uses the information set in the load.py file.
This file contains the Loader class, which reads the data and save it
to the database according to the model.
The loader for this pipeline would be:

\begin{verbatim}
class Loader(run.Loader):

    def setup(self):
        self.fp = open(settings.IRIS_PATH)

    def get_name_instance(self, row):
        name = self.session.query(
            models.Name
        ).filter(
            models.Name.name == row["Name"]
        ).first()

        if name is None:
            name = models.Name(name=row["Name"])

            # we need to add the new
            # instance and save it
            self.save(name)
            self.session.commit()

        return name

    def store_observation(self, row, name):
        return models.Flower(
            name=name,
            sepal_l=row["SepalLength"],
            sepal_w=row["SepalWidth"],
            petal_l=row["PetalLength"],
            petal_w=row["PetalWidth"])

    def generate(self):
        for row in csv.DictReader(self.fp):
            name = self.get_name_instance(row)
            obs = self.store_observation(
                row, name)
            yield obs

    def teardown(self, *args):
        if self.fp and not self.fp.closed:
            self.fp.close()
\end{verbatim}

In this example, \verb|setup| is executed just before \verb|generate|,
and it is the best place to open the data file.
On the other hand \verb|teardown| runs after \verb|generate| and uses
information about their error state.
The actual reading of each line in the data is split into two parts
within \verb|generate|:
The method named \verb|get_name_instance| receives the row as a
parameter and returns a \verb|IRIS.models.Name| instance referred to
the name of such file (Iris-virginica, Iris-versicolor, or
Iris-setosa).
Every time a name is non existant this method must create a new one
and store this model before returning it.
Another method, \verb|store_observation|, receives the row as a
parameter, and also the instance of IRIS.models.Name just created by
the previous model.
This method just needs to return the instance and deliver it to the
loader without saving it.
Finally, the \verb|yields obs| line within \verb|generate| put the
observation into the database through \textsc{corral} functionalities.

Once the data has been stored according to the model and is organized
into the database, all the processing steps can be written as
separate units into the \verb|steps.py| file.
Steps (and loaders) are controllers that do not need to run
sequentially.
Instead, a step will perform operations on
the available data when the pipeline is run.

The \verb|Statistics| class must be instantiated, for each different
name on the \verb|flowers| table, in the \verb|steps.py| file
using the definitions in \verb|models.py|:

\begin{verbatim}
from . import models

class StatisticsCreator(run.Step):

    model = models.Name
    conditions = []

    def process(self, name):
        stats = self.session.query(
            models.Statistics
        ).filter(
            models.Statistics.name_id == name.id
        ).first()
        if stats is None:
            yield models.Statistics(
                name_id=name.id,
                mean_sepal_l=0.,
                mean_sepal_w=0.,
                mean_petal_l=0.,
                mean_petal_w=0.)
\end{verbatim}

With the database ready to store the data, the mean can be computed
through a specific process for each Name in flowers.  For example,
for the ''setosa'' type:

\begin{verbatim}
class SetosaStatistics(run.Step):

    model = models.Statistics
    conditions = [
        models.Statistics.name.has(
            name="Iris-setosa"),
        models.Statistics.mean_sepal_l==0.]

    def process(self, stats):
        sepal_l, sepal_w = [], []
        petal_l, petal_w = [], []
        for obs in stats.name.flowers:
            sepal_l.append(obs.sepal_l)
            sepal_w.append(obs.sepal_w)
            petal_l.append(obs.petal_l)
            petal_w.append(obs.petal_w)
        stats.mean_sepal_l = sum(sepal_l)
        stats.mean_sepal_w = sum(sepal_w)
        stats.mean_petal_l = sum(petal_l)
        stats.mean_petal_w = sum(petal_w)
        stats.mean_sepal_l = (
            stats.mean.sepal_l/len(sepal_l))
        stats.mean_sepal_w = (
            stats.mean.sepal_w/len(sepal_w))
        stats.mean_petal_l = (
            stats.mean.petal_l/len(petal_l))
        stats.mean_petal_w = (
            stats.mean.petal_w/len(petal_w))
\end{verbatim}

Similarly, the corresponding classes must be written for the other two
Iris types.
For the corral framework to be aware of the steps, they must be added
in the \verb|settings.py| file, as mentioned previously.
Finally, the pipeline can be run using the following command line:

\begin{verbatim}
   $ python in_corral run
\end{verbatim}


\subsection{Processing of exoplanet data}
\label{appendix_exo}

The pipeline for the exoplanets has a different data structure, so its
definition can be, for example:

\begin{verbatim}
class Planet(db.Model):

    __tablename__ = 'Planet'

    id = db.Column(
        db.Integer, primary_key=True)

    nomb = db.Column(db.Float, nullable=False)
    per = db.Column(db.Float, nullable=False)
    mass = db.Column(db.Float, nullable=False)
    sep = db.Column(db.Float, nullable=False)
    dist = db.Column(db.Float, nullable=False)
    mstar = db.Column(db.Float, nullable=False)
    rstar = db.Column(db.Float, nullable=False)
    teff = db.Column(db.Float, nullable=False)
    fe = db.Column(db.Float, nullable=False)
\end{verbatim}

The load of the data is performed with just one table.
Then, the Loader class contain the setup and teardown functions, which
are the same than those in the IRIS example, except for the path to the
data file, given in this case by \verb|settings.EXO_PATH|.
The generate function is:

\begin{verbatim}
import csv
from corral import run
from corral.conf import settings
from exo import models

class Loader(run.Loader):
    """Extract data from the `exoplanets.csv` and feed
    the stream of the pipeline.

    """

    def setup(self):
        # we open the file and assign it to
        # an instance variable
        self.fp = open(settings.EXO_PATH)

    def float_or_none(self, value):
        try:
            return float(value)
        except (TypeError, ValueError):
            return None

    def generate(self):
        # now we make use of "self.fp"
        # for the reader
        for row in csv.DictReader(self.fp):
            di = {
                'name': row['NAME'],
                'per': self.float_or_none(
                    row['PER']),
                'mass': self.float_or_none(
                    row['MASS']),
                'sep': self.float_or_none(
                    row['SEP']),
                'dist': self.float_or_none(
                    row['DIST']),
                'mstar': self.float_or_none(
                    row['MSTAR']),
                'rstar': self.float_or_none(
                    row['RSTAR']),
                'teff': self.float_or_none(
                    row['TEFF']),
                'fe': self.float_or_none(
                    row['FE'])}
            yield models.Planet(**di)

    def teardown(self, *args):
        # checking that the
        # file is really open
        if self.fp and not self.fp.closed:
            self.fp.close()
\end{verbatim}
where the function Empty2None allows to deal with missing values, which
is common in exoplanet data.
This pipeline can also be extended by adding steps and alerts.
For instance, a step can be configured to filter the dataset, compute
correlation parameters, or apply machine learning techniques to
discover clustering or perform classifications.
If new planets are added to the data file, running the pipeline
updates all the results previously computed.
Also, the python environment allows to write alerts, which
can be configured to produce plots, send the results by
email or replace older versions in a webpage.
Here we show an example of a step, which determines the list of planets in the habitable zone, and of an alert, which performs a scatter plot of mass and period of planets in the habiltable zone.

In order to add the list of planets in the habitable zone to the database, we create a new table as follows:

\begin{verbatim}
class HabitableZone(run.Step):
   __tablename__ = "HabitableZoneStats"

    id = db.Column(
        db.Integer, primary_key=True)

    planet_id = db.Column(
        db.Integer, db.ForeignKey('Planet.id'),
        nullable=False)
    planet = db.relationship("Planet",
        backref=db.backref("hzones"))

    luminosity = db.Column(db.Float)
    radio_inner = db.Column(db.Float)
    radio_outer = db.Column(db.Float)

    in_habitable_zone = db.Column(db.Boolean)
\end{verbatim}

Then, the step performs the search of planets that fulfill the requirement of being in the habitable zone.
To that end, we compute the boundaries of the habitable zone as $rin=L/(1.1*L_{sun})$ and $rout=L/(0.53*L_{sun})$, and the luminosity as
\begin{equation}
L = 4 \pi R_{\ast}^2 * \sigma T_{eff}^4
\end{equation}
This is performed on the \verb|models.Planet| model for the planets
that have both the period and the mass measured
in the dataset.
All these planets that satisfy the condition of being in the habitable zone
are then ingested to the data base on the \verb|HabitableZoneStats| table.

\begin{verbatim}
from corral import run
import numpy as np
from astropy import units as u, constants as c
from . import models

STEFAN_BOLTZMANN = c.sigma_sb
SUN_LUMINOSITY = c.L_sun

class HabitableZone(run.Step):
    """Compute some statistics of the star of
    a given planet and then determines if is in
    their habitable zone.
    """

    model = models.Planet
    conditions = [model.rstar != None,
                  model.teff != None]

    def process(self, planet):
        # habitable zone of the host star
        Rstar  = (planet.rstar * u.solRad).to('m')
        Teff = planet.teff * u.K
        luminosity = (
            STEFAN_BOLTZMANN * 4 * np.pi *
            (Rstar ** 2) * (Teff ** 4))
        lratio = luminosity / SUN_LUMINOSITY
        rin = np.sqrt(lratio / 1.1)
        rout = np.sqrt(lratio / 0.53)

        in_hz = (
            planet.sep >= rin and
            planet.sep <= rout)
        return models.HabitableZoneStats(
            planet=planet,
            in_habitable_zone=in_hz,
            luminosity=lratio.value,
            radio_inner=rin.value,
            radio_outer=rout.value)
\end{verbatim}

Finally, we show an alert that produces a scatter plot of planet mass vs. period:

\begin{verbatim}
class LogScatter(ep.EndPoint):

    def __init__(self, path, xfield, yfield,
                 title, **kwargs):
        self.path = path
        self.xfield = xfield
        self.yfield = yfield
        self.title = title
        self.kwargs = kwargs
        self._x, self._y = [], []

    def process(self, hz):
        planet = hz.planet
        x = getattr(planet, self.xfield)
        y = getattr(planet, self.yfield)
        if x and y:
            self._x.append(x)
            self._y.append(y)

    def teardown(self, *args):
        plt.scatter(
            p.log(self._x),
            np.log(self._y), **self.kwargs)
        plt.title(self.title)
        plt.legend(loc="best")
        plt.savefig(self.path)
        super(LogScatter, self).teardown(*args)

class PlotAlert(run.Alert):
    """Store a list of planets in habitable
    zone in a log file and also generate a
    period vs mass plot of this planets
    """

    model = models.HabitableZoneStats
    conditions = [
        model.in_habitable_zone == True]
    alert_to = [
        ep.File("in_habzone.log"),
        LogScatter(
            "in_habzone.png",
            xfield="per", yfield="mass",
            title="Period Vs Mass",
            marker="*")]

    def render_alert(self, now, ep, hz):
        planet = hz.planet
        data = []
        for fn in planet.__table__.c:
            data.append([fn.name,
            getattr(planet, fn.name)])
        fields = ", ".join(
            "{}={}".format(k, v) for k, v in data)
        return "[{}] {}\n".format(
            now.isoformat(),  fields)
\end{verbatim}

In order to generate a report on the quality of this pipeline, we must perform some tests.
In what follows, we show two tests that check for consistency in the data.
The test in \verb|HabitableZoneTest| creates an instance of a planet with $R_{\ast}=1$ and $T_{eff}=1$.
This should produce a luminosity of $4*\pi*\sigma/L_{\odot}$, and should not be into the habitable zone.
The test also accounts for the consistency of the boundary values, checking that
the inner boundary is lesser than the outer boundary.
Finally, it verifies that as a result of the step, just one entry on the \verb|HabitableZoneStats| is produced.

\begin{verbatim}
class HabitableZoneTest(qa.TestCase):

    subject = steps.HabitableZone

    def setup(self):
        planet = models.Planet(
            name="foo", rstar=1, teff=1)
        self.save(planet)

    def validate(self):
        planet = self.session.query(
            models.Planet).first()
        hzone = planet.hzones[0]
        self.assertAlmostEquals(
            hzone.luminosity, 8.96223797571e-10)
        self.assertLess(
            hzone.radio_inner, hzone.radio_outer)
        self.assertFalse(hzone.in_habitable_zone)
        self.assertStreamCount(
            1, models.HabitableZoneStats)
\end{verbatim}

The other test checks that in the case a planet does not have the two values that are
required for the plot in the alert, no entry is generated on the \verb|HabitableZoneStats| table.

\begin{verbatim}
class HabitableZoneNoRstarNoTeffTest(qa.TestCase):

    subject = steps.HabitableZone

    def setup(self):
        planet = models.Planet(name="foo")
        self.save(planet)

    def validate(self):
        self.assertStreamCount(0,
             models.HabitableZoneStats)
\end{verbatim}

These tests produce a QA index of 27.31 per cent, and a qualification F with a 81.94 per cent coverage.
It must be noticed that in order to improve the quality of this pipeline, more tests should be prepared.
This example can be downloaded from the project
repository.
All the structural and quality documents can be found in GitHub pipeline repository
in the following directions:

\begin{itemize}
\item Pipeline repository, \url{https://github.com/toros-astro/corral_exoplanets}
\item QA report, \url{https://github.com/toros-astro/corral_exoplanets/blob/master/exo/qareport.md}
\item Model class diagram, \url{https://github.com/toros-astro/corral_exoplanets/blob/master/exo/models.png}
\item Pipeline documentation, \url{https://github.com/toros-astro/corral_exoplanets/blob/master/exo/doc.md}
\end{itemize}

\onecolumn

\newcommand{\specialcell}[2][c]{%
  \begin{tabular}[#1]{@{}c@{}}#2\end{tabular}}

\section{Comparative Table Between Several Pipeline Frameworks}
\label{appendixb}

In the following table we condense a collection of some of the most prominent pipeline implementations.
Given a particular problem situation and resource constrain,
it can be difficult to decide on a particular implementation of a pipeline
and it would be even harder to test every pipeline framework available.
See \autoref{section:comparison} for a discussion.

\begin{figure}[ht]
\centering
\includegraphics[height=.78\textheight,keepaspectratio=1]{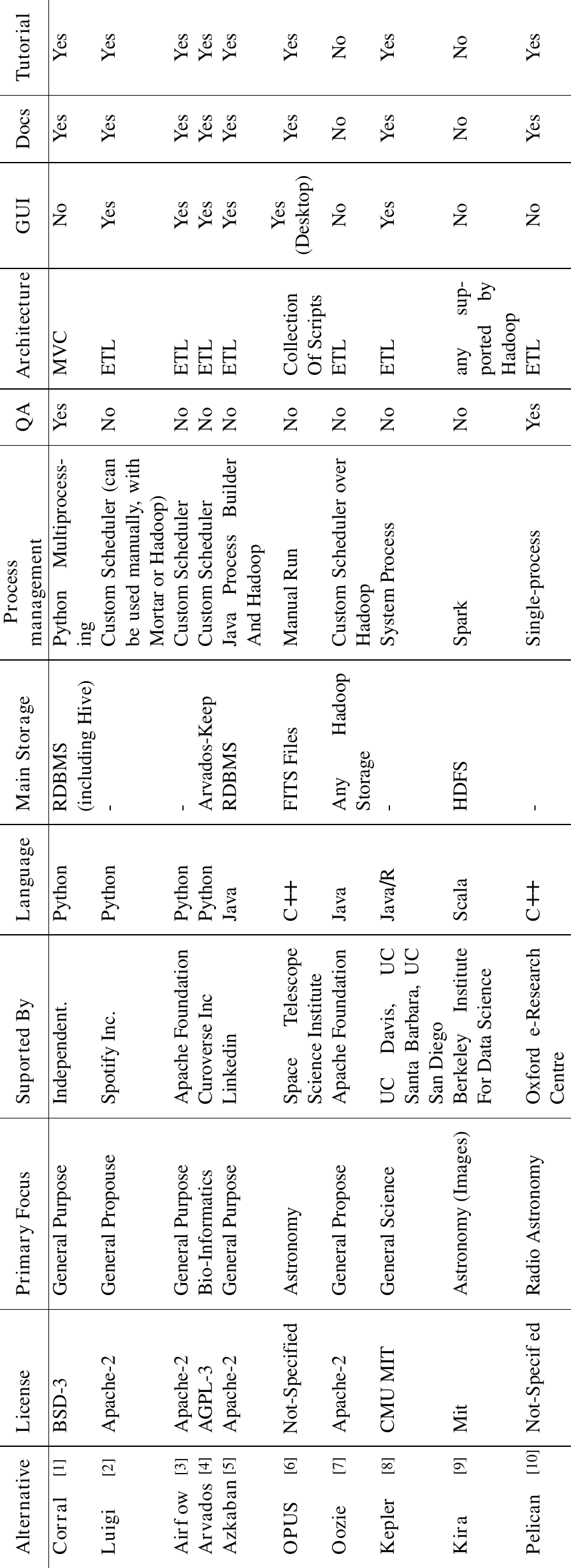}
\end{figure}

\label{fig:tabla}

\newpage
\raggedright
\textbf{Links}

{ \renewcommand\labelenumi{[\theenumi]}
\begin{enumerate}
\item \textbf{Corral:} \url{http://corral.readthedocs.io}
\item \textbf{Luigi:} \url{https://luigi.readthedocs.io}
\item \textbf{Airflow:} \url{http://airflow.apache.org/}
\item \textbf{Arvados:} \url{https://arvados.org}
\item \textbf{Azkaban:} \url{http://azkaban.github.io/azkaban}
\item \textbf{OPUS:} \url{http://www.stsci.edu/institute/software_hardware/opus/}
\item \textbf{Oozie:} \url{http://oozie.apache.org/}
\item \textbf{Kepler:} \url{https://kepler-project.org}
\item \textbf{Kira:} \url{https://github.com/BIDS/Kira/}
\item \textbf{Pelican:} \url{http://www.oerc.ox.ac.uk/~ska/pelican/}
\end{enumerate}
}

\end{document}